\documentclass[12pt]{article}
\usepackage{amsmath}
\usepackage{amssymb}
\usepackage{cite}
\usepackage{framed}
\usepackage{graphicx}
\usepackage{titletoc}
\usepackage{epsfig}
\usepackage{amsbsy}
\usepackage{amstext}
\usepackage{url}
\usepackage{subfig}
\usepackage{multicol}
\usepackage{mathtools, cuted}
\usepackage{adjustbox}
\begin{document}
	\title{Application of Vector Sensor for Underwater Acoustic Communications}
	\author{
$\mbox{Farheen Fauziya}^{1}$, $\mbox{Brejesh Lall}^{1,2}$ and $\mbox{Monika Agrawal}^{1,3}$ \\
				$^{1\#}$Bharti School of Telecomm. Techn. and Mgmt.\\
					$^{2\#}$Department of Electrical Engineering\\
					$^{3\#}$Centre for Applied Research in Electronics\\
					Indian Institute of Technology -- Delhi\\
					New Delhi, India 110016\\
				Emails:\small{ bsz148360@dbst.iitd.ac.in, brejesh@ee.iitd.ac.in, maggarwal@care.iitd.ac.in}	}
	\date{\today}
	\maketitle
	\begin{abstract}
		The use of vector sensors as receivers for Underwater Acoustic Communications systems is gaining popularity. It has become important to obtain performance measures for such communication systems to quantify their efficacy. The fundamental advantage of using a vector sensor as a receiver is that a single sensor is able to provide diversity gains offered by MIMO systems. In a recent work novel framework for evaluating capacity of underwater channel was proposed. The approach is based on modeling the channel as a set of paths along which the signal arrives at the receiver with different Angles of Arrival. In this work, we build on that framework to provide a bound on the achievable capacity of such a system. The analytical bounds have been compared against simulation results for a vector sensor based SIMO underwater communications system. The channel parameters are modeled by analysing the statistics generated with Bellhop simulation tool. This representation of the channel is flexible and allows for characterizing channels at different geographical locations and at different time instances. This characterization in terms of channel parameters enables the computing of the performance measure (channel capacity bound) for different geographical locations. 
	\end{abstract}
	\maketitle
	\section{Introduction}\label{Sec:1}

With the development of vector sensors for underwater acoustic communications there has arisen a need to obtain performance measures for such communications systems. The ability of a single vector sensor to provide diversity gains has been demonstrated in some recent works \cite{nehorai1994acoustic, Abdi_Guo09T}. But how effective is the diversity that a single vector sensor can provide? Researchers have made some attempts to answer this question by estimating the capacity of these communications system. This is however an arduous task since the underwater channel is very complex and difficult to model \cite{Stojanovic09,pbra014e_ch23}. A large volume of research with multicarrier transmission in the form of OFDM is done to overcome the inter-symbol interference resulting from the frequency selectivity of underwater acoustic communication channel \cite{Bai_IET,iet-com.2017.0037}. In \cite{iet-com.2017.0037}, orthogonal frequency division multiplexing (OFDM) is taken into account with cooperative transmission to improve the performance and to take advantage from the spatial diversity.
Besides, the channel is very topography specific i.e. has strong dependence on the location. A recent attempt has been made to develop an AoA based framework for representing the shallow underwater channel \cite{Fauziya_16}. This framework is simple and intuitive and can easily be adapted to represent different channel behaviors. In this work, we use that framework to obtain an upper bound on capacity of the underwater channel. A tight upper bound has been obtained in this work. 

Work on capacity computation of MIMO channels in underwater communications has been reported in \cite{Bouvet_10,Hicheri_14}. However, all these work consider MIMO implemented using an array of scalar sensors. Radosevic et.al. \cite{Radosevic_10} use probe signals to estimate the statistical properties of the time varying channel. They show that Rician fading is a good match for experimental data. This model is used to evaluate the channel capacity for both the Single Input Single Output (SISO) and MIMO systems. In another work \cite{Jin_Gao_07}, the authors obtain a lower bound on achievable information rate for MIMO communications over underwater acoustics channels. They evaluate the ergodic capacity for partial channel state information (CSI) at the transmitter and no CSI at the transmitter scenarios. That work also uses an array of scalar sensors to achieve MIMO capability in underwater acoustic communication systems. Little work is available on channel capacity computation for vector sensor based underwater MIMO communications. One work that does attempt this is by Abdi el.at. \cite{Guo_A12,Guo_chen12}. That work assumes that the number of transmitters is large. This is not a realistic assumption when one is attempting to compute the MIMO capacity of communications system, which consists of a single scalar transmitter and a single vector receiver. Besides, in that work different capacity expressions have been obtained for low SNR and high SNR scenarios. Finally, the normalization performed on the correlation matrix for the vector sensor case is not entirely justified. However, the work provides a first approach for computing the maximum data rates of the particle velocity channels. In \cite{Fauziya_16}, the authors present an AoA framework for computing the underwater channel capacity. We exploit that framework to obtain capacity bounds that are not restricted by the simplifying assumptions used in that work.

It is well known that acoustic shallow water channel is very complex and location specific. Numerous efforts have been made to come up with good propagation models which can be used for generating simulated data for different underwater acoustic channel scenarios. Some of the models are based on Ray theory, while others use model expansions and wave number techniques. Models are classified as either being range dependent or range independent. In a range dependent model, the environmental parameters  like water depth, sound speed etc are kept fixed with range whereas in range independent models, the parameters are allowed to vary with distance. Ray theory is based on simplification applied to the wave equation and can be viewed as a high frequency approximation of the same. The method is however reasonably accurate for communication over short and medium range. In models based on Ray theory, trajectories of rays from source are computed. Ray tracing method have the advantage of being fast, amenable to incorporating directionality and of being accurate. A very popular Ray theory based model is Bellhop \cite{Mic_Bellhop,Finn_Bellhop}, which is used in this work. Beam tracing is a variant of ray tracing \cite{Porter_Bucker}; it applies information about beam width associated with each ray to determine amplitude of the pressure, thus overcoming the shadow zone problems associated with the ray tracing methods. Bellhop predicts acoustic pressure fields in ocean environments and uses ray theory to provide an accurate deterministic picture of underwater acoustic channel for a given geometry and signal frequency. Bellhop provides the flexibility of simulating different environmental conditions and accept the corresponding information via the configuration file. The simulation parameter uses in this work are shown in Table \ref{bellhop_par}.

To derive the capacity, the following model of the channel has been assumed \cite{Fauziya_16}: The path gains are usually modeled as Rayleigh distributed  \cite{Win_06,Chitre_08} and we use the same model for path gain. The other aspect of shallow water channel is that the path gain is a function of the AoA. This dependence is analyzed using the Bellhop simulation tool and is incorporated into the channel model. A tight upper bound is computed based on this model of underwater channel. The specific contributions of this paper are listed as follows, \\
(i) AoA based framework  for channel characterization. 
\\(ii) The channel parameters models are obtained.
\\ (iii) Channel capacity bound is obtained.
\\ (iv) The bound is compared to SISO channel capacity.
This paper is organized as follows: This section is the
introduction. Sections \ref{Sec:2} presents brief introduction of the acoustic vector sensor, section \ref{Sec:2a} contains description of the system model. In section \ref{Sec:3} we
derive the capacity of vector sensor based SIMO system. Mathematical description and closed form expression for the capacity are presented.
Section \ref{Sec:4} contains the numerical results, and the conclusions are given
in section \ref{Sec:5}. For enhancing readability of the paper, the list of symbols is given in Table \ref{Symbol_Tab}.
\begin{table*}[h!]
	\caption{List of main symbols}
	\centering
	\begin{adjustbox}{width=1\textwidth}
		\small
		\begin{tabular}{l c c c}
			\hline \hline 
			$s$ & Transmitted Signal\\ 		
			$r$ & Scalar component of the received acoustic pressure signal \\
			$r^{y}$ & Pressure equivalent velocity component along range  \\
			$r^{z}$ & Pressure equivalent velocity component along depth  \\
			$n$ & Ambient noise pressure \\
			$n^{y}$ & Pressure equivalent ambient noise velocity component along range  \\
			$n^{z}$ & Pressure equivalent ambient noise velocity component along range  \\
			$h$ & Channel impulse response of scalar component of vector sensor  \\
			$h^{y}$ & Channel impulse response of particle velocity component along range  \\
			$h^{z}$ & Channel impulse response of particle velocity component along depth  \\
			$\gamma, \gamma_{ib}, \gamma_{is}$ & Angle of arrival of reflected rays, $i^{th}$ rays reflected from the bottom and $i^{th}$ rays reflected from the surface\\
			$h_{LoS}, h_{ib}, h_{is}$ & Path gain of LoS path, $i^{th}$ path from bottom and $i^{th}$ path from surface \\
			$\tau_{LoS}, \tau_{ib}, \tau_{is}$ & Path delay of LoS path, $i^{th}$ path from surface and $i^{th}$ path from bottom \\
			$\Lambda, \xi, \varsigma$ & Scaling parameter, Mean value and Spreading factor of scaled Gaussian function \\
			$\theta_{ib}, \theta_{is}, \beta_{ib}, \beta_{is} $ & Parameters of the AoA pdf \\
			$N_{c}$ & Noise covariance matrix at receiver \\
			$E[.]~ *~ \delta(.)$ & Expectation operator, Convolution operator and Dirac delta function respectively\\
			$(.)^T$~ \& $(.)^\dagger$ & Transpose and Hermitian operators respectively \\
			\hline \hline
		\end{tabular}
	\end{adjustbox}
	\label{Symbol_Tab}
\end{table*}
\section{Acoustic Vector Sensor}\label{Sec:2}
Before describing the system model, we present a brief discription of acoustic vector sensor. An acoustic vector sensor is different from a scalar acoustic sensor, in that it measures both the acoustic pressure and the acoustic velocity. This information can be used to estimate the acoustic intensity vector, which is a measure of the acoustic energy flow rate per unit area in the direction perpendicular to the flow. Mathematically acoustic intensity vector can be represented a product of acoustic pressure and acoustic particle velocity. Vector sensors are usually designed in one of the following two ways: (i) Pressure-velocity based method resulting in a class of vector sensors called inertial sensors and (ii) Pressure-pressure based method leading to a class of vector sensors categorized as gradient sensors.

Inertial sensors work directly on the particle velocity. The particle velocity can be obtained either by time integration of the accelerometer measurements or by direct measurement of velocity e.g. using a Microflown. Inertial sensors offer the advantage of broad dynamic range, however, packaging of the sensor without affecting its response to the motion is a challenge. Also, such sensors do not distinguish between acoustic waves and non-acoustic motion (e.g. support structure vibrations) and must, therefore, be properly shielded from such disturbances.

In gradient sensors, finite difference approximation of the spatial derivative of the sound-field is used to compute the particle velocity. Such sensor require that the separation d between the two microphones should be much smaller than the acoustical wavelength, $\lambda$. An increase in separation (for a fixed frequency) leads to an increase in error in the pressure-gradient estimate. Also, such sensors are sensitive to sensor noise, microphone mismatch, reflection and diffraction of the source radiated signal, placement of the microphones, the structure that holds the microphones and the polar pattern characteristics of each microphone. A typical gradient sensor has closely spaced omni-directional and/or gradient-microphones that measure the pressure and pressure-gradients in the three orthogonal directions which are, in turn, used for particle velocity estimation. In summary, gradient sensors can be manufactured in smaller sizes and thus are more suitable for high frequency applications. However, the finite-difference approximation limits their operating dynamic range. Given their individual advantages and disadvantages, the choice of type of vector sensor depends on the application. 

The output of an AVS consists of four channels, namely, the acoustic pressure and the three Cartesian components of the particle velocity. In this paper the vector sensor considered gives a measure of three components, the acoustic pressure and two Cartesian components of particle velocity (those along range and depth).

\section{System Model}\label{Sec:2a}
\begin{figure}[t!]
	\centering
	\includegraphics[width=4.8in,height=3.in]{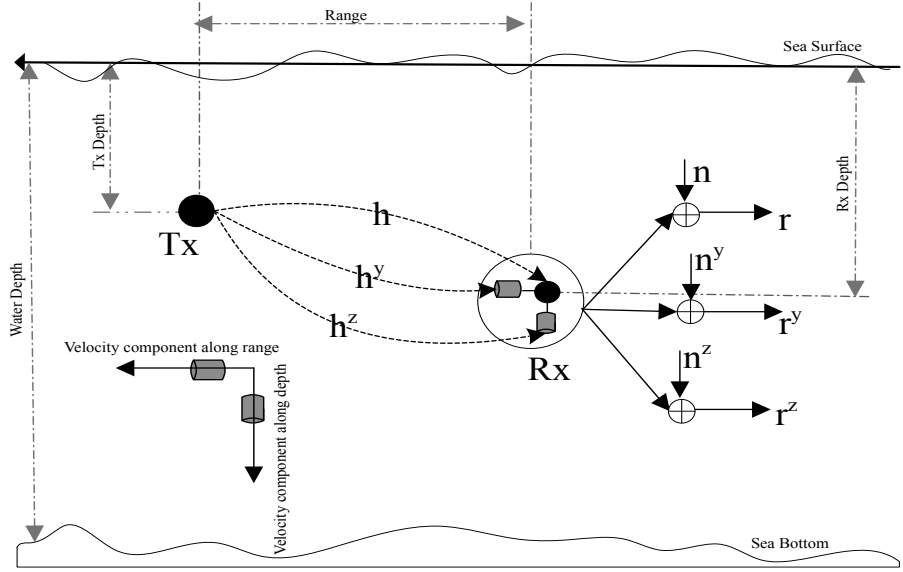}
	\vspace{-0.5em}
	\caption{\small Acoustic vector sensor receiver in underwater communications system.}
	\label{Fig:1}
\end{figure}

In this paper, we consider an underwater communications system with a scalar transmitter and a vector sensor receiver. The system model is shown in Fig.~\ref{Fig:1}.
The received signal consists of a scalar component and two vector components \cite{Abdi_Guo_T09}. To aid in understanding the representation proposed for the pressure equivalent velocity component along range and depth, a brief description of the mathematical model for acoustic flow is now presented. The relation between pressure and particle velocity can be represented as, 
\begin{align}
\bigtriangledown h=\rho \left\lbrace  (\vec{v}. \bigtriangledown) \vec{v} +\frac{\partial \vec{v}}{\partial t}\right\rbrace ,
\label{Eqn1}
\end{align}
where, $\vec{v}$ is the particle velocity, $h$ is the pressure and $\bigtriangledown$ is the Laplace operator. Using the linearized momentum align the relation between acoustic particle velocity and pressure is given as,  \cite{KFCS_98},
\begin{align}
\bigtriangledown h= - \rho \frac{\partial \vec{v}}{\partial t}
\label{Eqn2}
\end{align} 

Using Eqn.~(\ref{Eqn2}), the velocity component along $ y ~\&~ z $ at frequency $f$ can be represented as follows,
\begin{align}
v^{y}=-\frac{1}{j \rho \omega}\frac{\partial h}{\partial y}, ~ v^{z}=-\frac{1}{j \rho \omega}\frac{\partial h}{\partial z}
\label{Eqn3}
\end{align}
where, $\rho$ is the fluid density, $\omega=2 \pi f$ and $j^{2}=-1$. From Eqn.~(\ref{Eqn3}), it is clear that velocity in a certain direction is directly proportional to the spatial pressure gradient in that direction. The acoustic pressure channel represents the scalar channel $h$ and the two particle velocity components represent the pressure equivalent velocity channels along range and depth denoted by $h^y$ and $h^z$ respectively.
The pressure equivalent velocity component along range and depth can be obtained by multiplying velocity channel $v^{y}~ \&~ v^{z}$ with the negative of acoustic impedance ($- \rho c$), where $c$ is the speed of sound. Incorporating the wave number $k=\omega/c=2 \pi/\lambda$ into the sound, the final expression for pressure equivalent velocity channel can be written as \cite{Abdi_Guo_T09},
\begin{align}
h^{y}=-\rho c v^{y}=\frac{1}{j k}\frac{\partial h}{\partial y}, ~ h^{z}=-\rho c v^{z}=\frac{1}{j k}\frac{\partial h}{\partial z},
\label{Eqn4}
\end{align}

\begin{figure}[h!]
\centering
\includegraphics[width=4.8in,height=3.in]{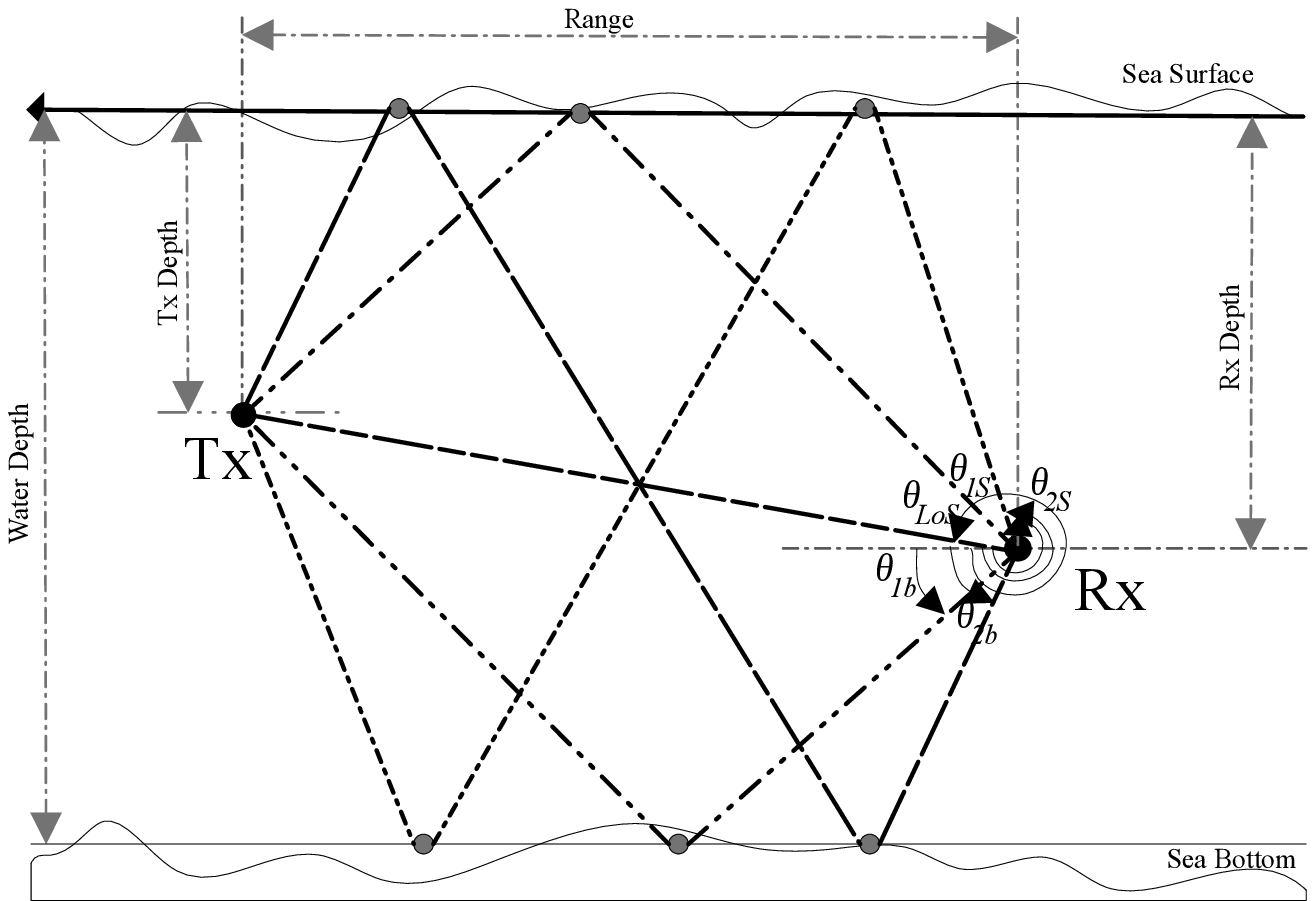}
\caption{System model based on vector sensor receiver}
\vspace*{-0.5em}
\label{Fig:2}
\vspace{-1.2em}
\end{figure}
The received signal components can be viewed as convolution of the transmitted signal with the three channel impulse responses. The pressure component of the received signal can be represented as follows,
\begin{align}
r=s*h+n.
\label{Eqn:5} 
\end{align}
Similarly received signal components along range and depth can be written as
\begin{align}
r^{y}=s*h^{y}+n^{y}, \quad r^{z}=s*h^{z}+n^{z} ,
\label{Eqn:6}
\end{align}  

where, $r$ and $s$ represent received and transmitted signals respectively, $h^y$ is the pressure equivalent velocity component along the range and $h^z$ is the pressure equivalent velocity component along the depth.
$r^{y}$ is received signal component along the $y$ direction (range), $r^{z}$ is received signal component along $z$ direction (depth),   
$n, n^y$ and $n^z$ are the ambient noise pressure and pressure equivalent ambient noise velocities along range and depth respectively. This communication systems can be represented as a $1\times3$ SIMO. The channel matrix $H$ is given as follows.
	
	\begin{equation}
	H=
	\begin{bmatrix}
	h\\
	h^{y}\\
	h^{z}
	\end{bmatrix}
	\label{Eqn:7}
	\end{equation}
	
	\begin{figure}[h!]
		\centering
		\includegraphics[width=4.2in,height=3.in]{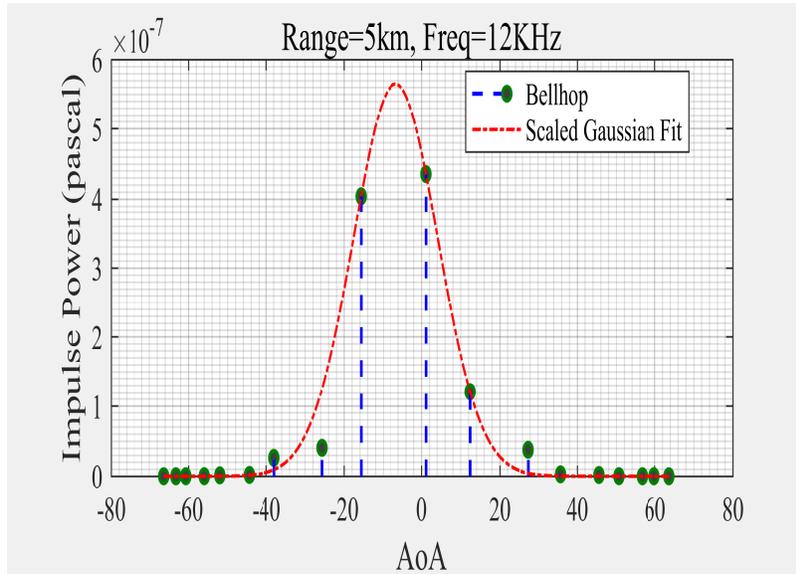}
		\vspace{1.02em}
		\caption{\label{Fig:3}{Dependence of path gain on the Angle of Arrival of the path}}
		\vspace*{1em}
	\end{figure}
	Given this system model, we now present brief a description of the statistical properties of the channel and how they have been incorporated in the channel model.
	In this work, it has been assumed that the ray will scatter only from the surface and bottom of the sea. The path gain has been assumed to be Rayliegh distributed and the sacale parameter of the density function depends on the AoA.
	
	\begin{equation}
	p_{h|\gamma} (\alpha|\gamma)= \frac{\alpha}{\sigma^2(\gamma)}
	\exp\Big(-\frac{\alpha^{2}}{2\sigma^2(\gamma)}\Big)~~; ~~~~~~~~h \ge 0
	\label{Eqn:8}
	\end{equation}
	
	where, $\gamma$ is the AoA and $\sigma$ is the scale parameter. At the receiver side, amplitude will be maximum if the ray comes along the direct path (i.e. with AoA $\theta_{LoS}$) and as one moves away from LoS the amplitude in either direction will decrease till it reaches zero at angles $\pm \pi/2$. To account for this variation in amplitude as a function of AoA, the scale parameter of Rayleigh distribution $\sigma^2$ is  modeled as a (scaled Gaussian) function of AoA with maximum amplitude at $\theta_{LoS}$. No studies exist on the distribution of path gain as a function of AoA. However, studies on distribution of path gain as a function of time of arrival show the dependence as Gaussian \cite{Qarabaqi2009StatisticalMO, Qarabaqi_13}. 
	Drawing inspiration from those works and by performing statistical analysis using data generated using Bellhop \cite{Mic_Bellhop}, we found that the best fit for the dependence of path gain on AoA is scaled Gaussian. Given that, the energy corresponding to a Rayliegh distribution is directly proportional to $\sigma^2$ we choose the map between AoA and Rayleigh distribution parameter to be scaled Gaussian.
	The analysis reveals that the best fit for the path gain statistics (as a function of AoA) is scaled Gaussian,
	\begin{equation}
	\sigma^{2}(\gamma)=\Lambda e^{-\!\left(\frac{\gamma-\xi}{\varsigma}\right)^2}
	\label{Eqn:9}
	\end{equation}
	
	where, $\Lambda, ~ \xi, \& ~ \varsigma$ corresponds to scale, mean and spreading factor of the fitted distribution. To illustrate the correctness of the fit, the comparison between the simulation statistics and the scaled Gaussian curve used to fit those statistics is shown in Fig.~\ref{Fig:3}. The figure clearly shows that scaled Gaussian is a very good approximation of the path loss statistics as a function of AoA. The goodness of fit has been tested using some popular measures and the resutls reveal a good fit \cite{Qarabaqi_13}. The values for SSE, R-square and RMSE respectively are $8.08e^{-08}, 0.9334$ and $7.215e^{-05}$. These results are for a fit with $95\%$ confidence bound, which clearly show that the fit is good \cite{Qarabaqi_13}.
	Another point to note is that this model for dependence of path gain on AoA implictly incorporates the effect of bottom reflection coefficients. The bottom reflection coefficients are a function of the grazing angle \cite{Hawkes_03, Rado_09}. Smaller angles result in higher attenuation and larger angles lead to lower attenuation. The paths that arrive with high AoA constitute rays that have grazed the bottom at lower angles and hence have suffered from greater attenuation. This is reflected in the relatively steep reduction in path gain as a function of AoA. If this dependence had been ignored the path gain would have been proportional to the cosine of the AoA, however by including the effect of bottom reflection coefficients the appropriate fit for the path gain is scaled Gaussian and not a cosine map. This is borne out by the curve fitting analysis.
	
	\begin{figure}[h!]
		\centering
		\includegraphics[width=4.5in,height=2.9in]{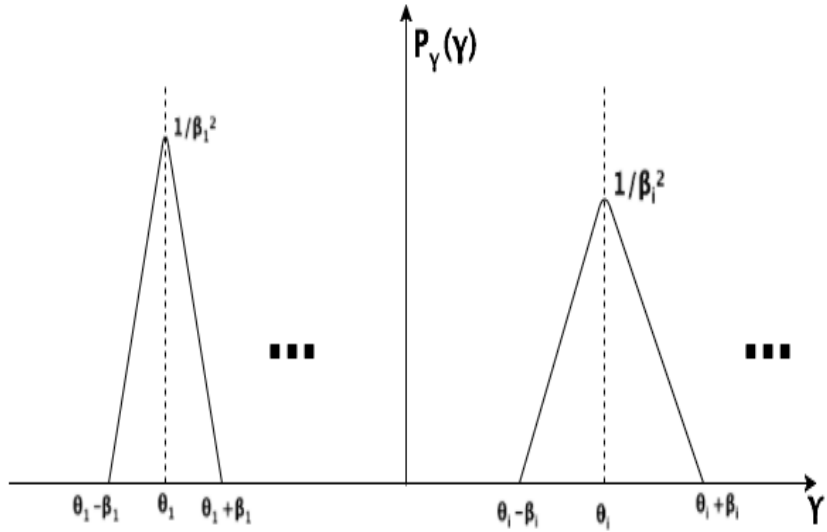}
		\vspace{1.02em}
		\caption{\label{Fig:4}{Probabilty density function characterizing the AoA distribution of the $i^{th}$ path.}}
		\vspace*{-1em}
	\end{figure}
	The other characteristic of the channel that one needs to model is the probability distribution of the AoA, $\gamma$ of the received path. In our work, the probability density function is represented using the following equations,
	\begin{equation}
	p_{\gamma_{i}}(\gamma_{i})=
	\begin{cases}
	\!\frac{1}{\beta^2_{i}}\left(\theta_{i}+\beta_{i}-\gamma_{i}\right)~; ~~~~~~ \theta_{i} <\gamma_{i}<\theta_{i}+\beta_{i} \\
	\!\frac{1}{\beta^2_{i}}\left(\gamma_{i}-\theta_{i}+\beta_{i}\right)~; ~~~~~~ \theta_{i}-\beta_{i} <\gamma_{i}<\theta_{i},
	\end{cases}
	\label{Eqn:10}
	\end{equation}
	where, $\gamma_{i}$ is the AoA of rays, $\theta_{i}$, is the parameters of the probability density function characterizing the AoA of the $i^{th}$ path. $\beta_{i}$ is the spread of AoA about $\theta_{i}$.
	In literature the AoA distributions have been variously modeled. In their work Abdi et.al. \cite{Abdi_Guo_T09} use Gaussian with very small variance to model the AoA distribution. However, no work categorically assigns a distribution to the AoA. In wireless communications, researchers have proposed Laplacian as a distribution of choice \cite{Suraw_05}. There is however consensus on the fact that the distribution of the AoA for a path is of a small variance \cite{Abdi_Guo09T,Abdi_Guo_T09}. Further, it can be shown that Gaussian (as also Laplacian) with small variance can be modeled using triangular distribtution function\cite{Weng1997}. For this reason and for aiding mathematical tractability we approximate the density funtion of the AoA as triangular \cite{Weng1997}.
	
	Given these models for the channel characteristics, we continue our description of the underwater acoustic communications system. The channel response can be represented in terms of individual path gains and AoA's as follows:
	\begin{equation}
	h=\sum_{i=1}^N h_i\delta(\gamma-\gamma_i)\delta(\tau-\tau_i)\delta(\gamma) ,
	\label{Eqn:11}
	\end{equation}
	where $N$ is the total number of paths, and  as shown in Fig.~\ref{Fig:2}, $h_{i}$ is the path loss along path $i$ and $\gamma_i$ is the AoA of the $i^{th}$ path. Using the framework introduced in this section,
	\begin{equation}
	h^y_i=\sum_{i=1}^Nh_i \cos[\gamma_i]\delta(\gamma)\delta(\tau-\tau_i) ,
	\label{Eqn:12}
	\end{equation}
	represents the path gain along the range. Similarly path gain along depth can be represented as
	\begin{equation}
	h^z_i=\sum_{i=1}^Nh_i \sin[\gamma_i] \delta(\gamma-\pi/2)\delta(\tau-\tau_i),
	\label{Eqn:13}
	\end{equation}
	
	where, $N$ is the number of paths reaching the receiver after no/single/multiple reflection/s from the surface and the bottom. For this framework, the probability density function of the AoA is represented as triangular and the path loss is modeled using a family of Rayleigh pdfs \textbf{\cite{Rado_09}}. The variance of Rayleigh pdfs is a function of AoA of that path. This dependence is modeled using a scaled Gaussian function with the max value at $\theta_{LoS}$ (the angle that LoS path makes with respect to range).
	Simulation parameters for Bellhop used in this analysis are listed in Table \ref{bellhop_par}.
	
	\subsection{AoA Representation}
	\begin{figure}[h!]
		\centering
		\includegraphics[width=\linewidth]{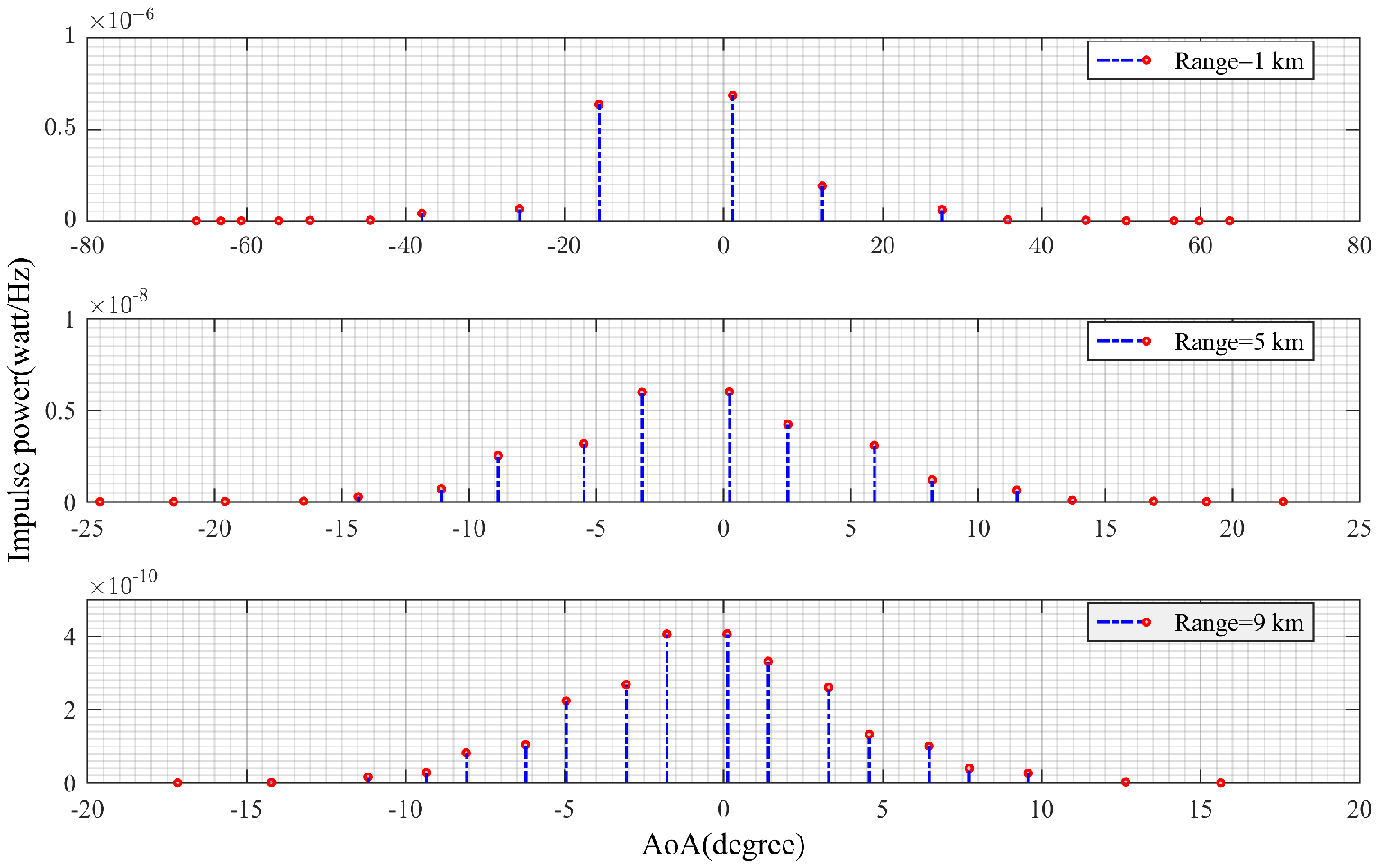}\\
		\includegraphics[width=\linewidth]{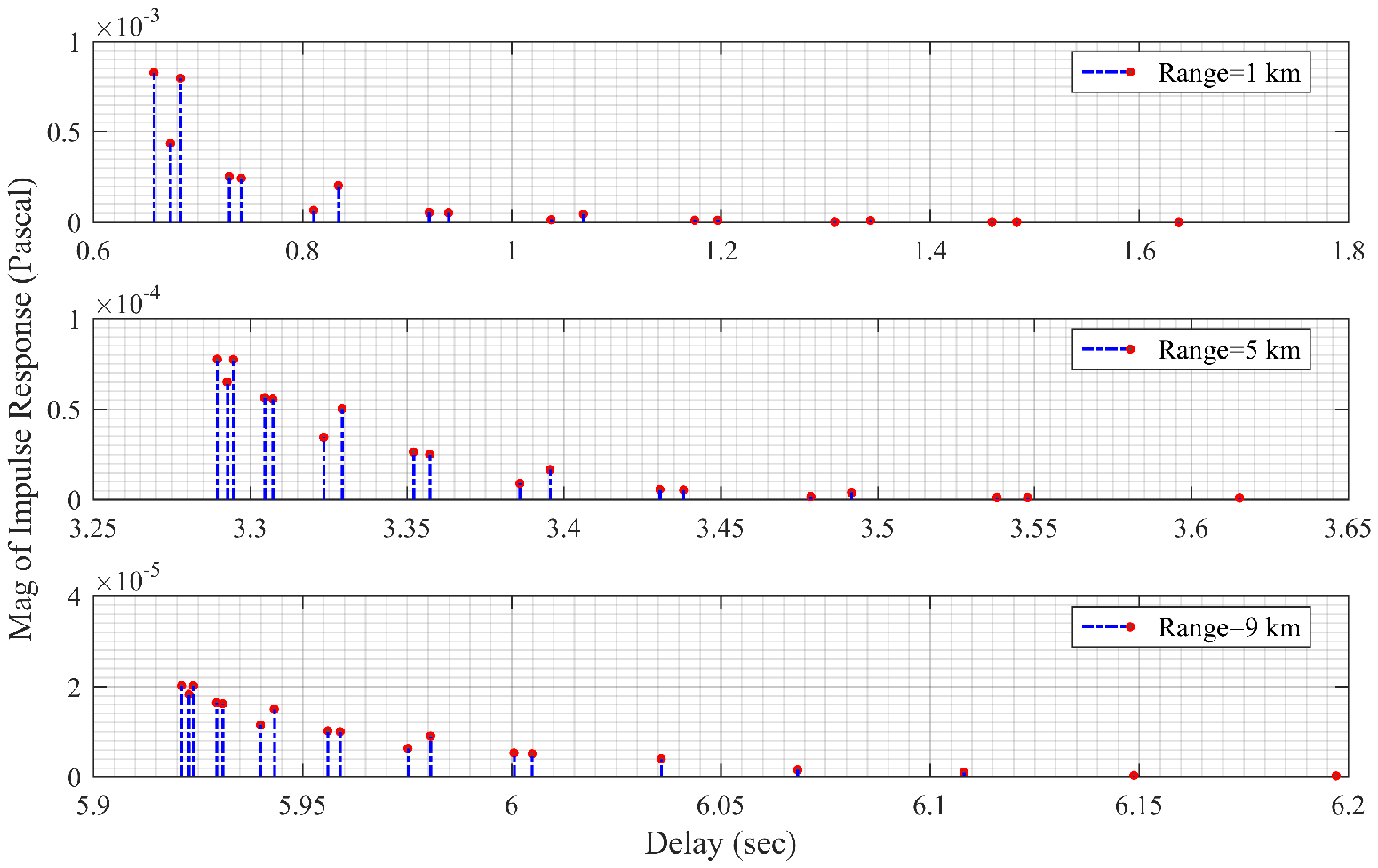}
		\caption{\label{Fig:5}{Impulse response as a function of (a) Delay and (b) AoA for different ranges}}
		\vspace*{1em}
	\end{figure}
	The propagation speed varies with water depth and hence in deep water the acoustic rays bend towards the layer with small propagation speed. However, in shallow water environment, the propagation occurs with constant speed. Therefore, the shallow water environment is considered to be an isovelocity medium (i.e. no refractions only reflections) and acoustic rays travel along straight lines. However, due to severe multipath effect caused by surface and bottom reflections, the transmitted signal propagates over large number of paths in shallow water. The two dimensional channel model for shallow water environment is bounded by oceans surface and bottom as shown in Fig.~\ref{Fig:2}.
	
	The received signal in a typical shallow underwater acoustic communications scenario consist of multiple paths arriving at different time instances. Given the relatively small speed of an acoustic wave in water and relatively large range the arrival times are typically displaced by several symbol durations. Accordingly, the channel response can be represented as Eqn.(\ref{Eqn:11}).
	In the shallow water scenario, the multiple paths can be attributed to reflections from the surface or bottom or from both. These reflections imply that unique AoA is associated with every individual path.
	
	\subsection{Channel Modeling}
	Given the channel representation in Eqn.~(\ref{Eqn:11}), Eqn.~(\ref{Eqn:12}) $\&$ Eqn.~(\ref{Eqn:13}) we now need to apply the model for AoA, $\gamma$, and compute the corresponding path gain, $h$, to fully characterize the channel. In literature, the path gain in UWAC systems has been modeled as Rayleigh distributed \cite{Hicheri_14}. However, in UWAC systems the individual paths arrive along multiple angles and at multiple time instances. Since the path gain depends on number of reflections and hence the AoA, this dependence can be incorporated into the distribution as follows,
	\begin{equation}
	p_{hi}(\alpha)=\frac{\alpha}{\sigma_{i}^{2}(\gamma_{i})}\exp(-\frac{\alpha^2}{2 \sigma_{i}^{2}(\gamma_{i})}) \quad \alpha \ge 0
	\label{Eqn:14}
	\end{equation}
	As mentioned earlier, the dependence of scale parameter on the AoA, has been obtained using statistical analysis of the path gain. Bellhop has been used to generate path gain and corresponding AoA data. The analysis reveals that the best map for representing the dependence of the scale parameter on AoA is scaled Gaussian.
	
	\begin{equation}
	\sigma^{2}(\gamma)=\Lambda e^{-\!\left(\frac{\gamma-\xi}{\varsigma}\right)^2}
	\label{Eqn:15}
	\end{equation}
	
	The value of $\Lambda, \xi~\& ~\varsigma$ can be heuristically estimated for the particular shallow water channel where the model is applied. The values used in this work are given in the Table \ref{bellhop_par}.
	
	The other characteristic of the channel that has been modeled is the probability density function of the AoA. In literature \cite{Rejeb2014}, the AoA has been variously modeled. Some popular representations include, Von Mises, Laplacian and even Gaussian distribution.
	\begin{equation} 
	p_{\gamma}(\gamma; \mu, \sigma)=\begin{cases}
	\frac{ A e^{-(\gamma-\mu)^2 /2 \sigma^2}}{\sigma \sqrt{2 \pi}} \quad - \pi/2 +\mu \le \gamma \le \pi/2 +\mu,\\
	0 \quad \text{otherwise}, 
	\end{cases}
	\label{Eqn:16}
	\end{equation}
	
	\begin{equation} 
	p_{\gamma}(\gamma; \mu, \sigma)=\begin{cases}
	\frac{ A e^{- \sqrt{2 }|\gamma-\mu|^2 / \sigma}}{\sigma \sqrt{2}} \quad - \pi/2 +\mu \le \gamma \le \pi/2 +\mu,\\
	0 \quad \text{otherwise},
	\end{cases}
	\label{Eqn:17}
	\end{equation}
	where, Eqn.~(\ref{Eqn:16}) represents the pdf of Gaussian distribution and Eqn.~(\ref{Eqn:17}) represents the Laplacian distribution. The  parameter $\sigma$ controls  the  spread  of  the 
	functions, while the constant $A$ is set such that area under the curve is one. For shallow water UWAC systems and for typical ranges the AoA are small and their spread about the mean value is smaller still. This observation allows us to use simplified pdf to make the performance analysis mathematically tractable. The approximation given in Eqn.~(\ref{Eqn:10}) results in negligible error, however it makes the analysis mathematically tractable \cite{Keith_AoA_08}.
	
	Given this channel characterization, we now derive the channel capacity in the following sub-section.
	\section{Channel Capacity}\label{Sec:3}
	In the previous section, we have described the channel and proposed an AoA based model for it. In this section, we compute the channel capacity using the proposed channel characterization. Channel capacity can be computed in terms of the mutual information between the transmitted and received signal.
	The mutual information between the transmitted and received signals is defined as,
	
	\begin{equation}
	I(S;R)=\sum_{s \epsilon A}\sum_{r \epsilon B}p(s,r)\log\left( \frac{p(s,r)}{p(s)p(r)} \right) 
	\label{Eqn:18}
	\end{equation}
	and formally  the capacity is defined as,
	\begin{equation*}
	C=\underset{p_S(s)}{sup}I(s;r)
	\end{equation*}
	where, $sup$ is supremum.
	
	For the SIMO case with \textit{L} received antennas, the received signal at the $i^{th}$ antenna is,
	\begin{equation}
	r^{(i)}(n)=h^{(i)}(n)*s(n)+\eta^{(i)}(n) \quad \text{for}~~ i=1~to ~ L
	\label{Eqn:19}
	\end{equation}
	where, $h^{(i)}(n)$ is the impulse response corresponding to the $i^{th}$ subchannel. Here, $s(n)$ is the transmitted signal and $\eta^{(i)}(n)$ is the noise at the $i^{th}$ receive antenna.
	
	For the current setting $L=3$, corresponding to the scalar sensor and the two components of the vector sensor, the corresponding signals are,
	\begin{align}
	\label{Eqn:20}
	&r=h(n)*s(n)+\eta(n) \nonumber \\
	& r^{(y)}=h^{y}(n)*s(n)+\eta^{y}(n)   \\
	& r^{(z)}=h^{z}(n)*s(n)+\eta^{z}(n) \nonumber
	\end{align}
	From \cite{Abdi_Guo_T09}, we know that the noise energy corresponding to the three diverse receive paths is $\Omega_{N}$, $\frac{\Omega_{N}}{2}$ and $\frac{\Omega_{N}}{2}$ respectively. Where, $\Omega_{N}$ is the variance of the additive Gaussian noise $\bold{\eta}$. Also these noise components are independent of each other. The impulse responses $h(n)$ can be represented in the terms of a vector $\bold{h}$, with each entry corresponding to the gain associated with the different paths of the sub-channels.
	
	The channel capacity will now be computed by applying the channel model characterization to Eqn.(\ref{Eqn:20}) 
	The ergodic capacity of MIMO communications system is given as follows \cite{Telatar_99,Bouvet_10}:
	\begin{equation}
	C=E\left[\log_{2} \left(\det(I+ \frac{\rho}{N_{t}} H^{\dagger}H)\right)\right]  bits/s/Hz,
	\label{Eqn:21}
	\end{equation}
	where,  $I$ is $N_{t}\times N_{t}$ identity matrix, $\rho$ is the SNR at the receiver and $H$ is the $N_{r}\times N_{t}$ channel matrix. $N_{t}$ is the number of transmitters and $N_{r}$ is the number of receivers. For the SIMO case, the ergodic capacity reduces to the following forms,
	\begin{equation}
	C=E\left[\log_{2} \left(\det(I+ \rho \sum_{i=1}^{L}\vert \bold{h}^{(i)}\vert\right)\right]  bits/s/Hz,
	\label{Eqn:22}
	\end{equation}
	For the vector sensor based SIMO system under consideration the ergodic capacity is given as follows,
	\begin{align}
	C= E \left[\log_{2}\left[1+\rho E \left(\vert \bold{h}\vert^{2}+2 \vert \bold{h^{y}}\vert^{2}+ 2 \vert \bold{h^{z}}\vert^{2}\right)  \right] \right],
	\label{Eqn:23}
	\end{align}
	were, $\bold{h},~ \bold{h^{y}} ~ \& ~ \bold{h^{z}}$ components are multiplied by two because the noise along these components is half of the noise along the corresponding scalar components. Also, as mentioned previously the three noise components are uncorrelated\cite{Abdi_Guo_T09}.
	Obtaining closed form for Eqn.~(\ref{Eqn:22}) is non trivial \cite{Paulraj_02}. In this paper, we obtain a bound on capacity and show how the bound compares to the simulated capacity.
	\subsection{Bound on Capacity :}
	As discussed above obtaining closed form expression for capacity is an arduous task. So in this subsection we obtain an upper bound  on capacity.
	Given the system representation, we now derive the channel capacity for this SIMO system. 
	%
	%
	
	To overcome the difficulties in obtaining a closed form for the above equation, we use Jensen's inequality to obtain the upper bound on capacity. Jensen's inequality \cite{vu2005characterizing} states that given a random variable $x$ and a convex function $\phi$, 
	\begin{equation*}
	E \left\lbrace  \phi(x) \right\rbrace   \le  \phi E\left\lbrace x \right\rbrace .
	\label{Eqn:24}
	\end{equation*}
	Noting that $\log_{2}\left[. ~\right] $ is a convex function, the channel capacity bound can be represented as follows,
	\begin{equation}
	\begin{split}
	C=&E\left[\log_2\left\lbrace 1+\rho (\left|\bold{h}\right|^2+2\left|\bold{h^{y}}\right|^2+2\left|\bold{h^{z}}\right|^2)\right\rbrace \right]   \\
	& \le \log_2\left[E\left\lbrace 1+\rho (\left|\bold{h}\right|^2+2\left|\bold{h^{y}}\right|^2+2\left|\bold{h^{z}}\right|^2)\right\rbrace \right].
	\end{split} 
	\label{Eqn:25}
	\end{equation}
	therefore,
	\begin{equation}
	\begin{split}
	C_{UB}=\left[\log_2\left\lbrace 1+\rho E (\left|\bold{h}\right|^2+ 2\left|\bold{h^{y}}\right|^2+2 \left|\bold{h^{z}}\right|^2)\right\rbrace \right]  ].
	\end{split} 
	\label{Eqn:26}
	\end{equation}
	The bound has been computed using system model shown in Fig.~\ref{Fig:2}. Here $\bold{h}$, $\bold{h^{y}}$ and $\bold{h^{z}}$ can be correlated. To handle this correlation, note that,

		\begin{align*}
		h=&\sum_{i=1}^{N}h_{i}\delta(\gamma-\gamma_{i})\delta(\tau -\tau_{i}),  \quad  h^{y}=\sum_{i=1}^{N}h_{i}\cos(\gamma_{i})\delta(\gamma)\delta(\tau -\tau_{i}), 
		\quad \\ &
		h^{z}=\sum_{i=1}^{N}h_{i}\sin(\gamma_{i})\delta(\gamma -\pi/2)\delta(\tau -\tau_{i})
		\end{align*}
	
	Now,\\

		$\vert \bold{h}\vert^{2}=\bold{hh^{\dagger}}=\sum_{i=1}^{N}\vert h_{i}\vert^{2}$,  $\vert \bold{h^{y}}\vert^{2}=\bold{h^{y}(h^{y})^{\dagger}}=\sum_{i=1}^{N} \vert h_{i}\vert^{2}\cos^{2}(\gamma_{i})$ and $\vert \bold{h^{z}}\vert^{2}=\bold{h^{z}(h^{z})^{\dagger}}=\sum_{i=1}^{N}\vert h_{i}\vert^{2}\sin^{2}(\gamma_{i})$ 

	Incorporating this in Eqn.(\ref{Eqn:26}), we obtain
	\begin{align}
	C_{UB}&=\log_{2}\left[1+\rho E_{h,\gamma} \left(\sum_{i=1}^{N}\left(\vert h_{i}\vert^{2}\left(1+2\cos^{2}(\gamma_{i})+2\sin^{2}(\gamma_{i})\right) \right)\right) \right]  \nonumber \\
	& =\log_{2}\left[1+3\rho E_{h,\gamma} \left(\sum_{i=1}^{N}\left(\vert h_{i}\vert^{2}\right)\right) \right] \nonumber \\
	& =\log_{2}\left[1+3\rho\int \int \sum_{i=1}^{N}\vert h_{i}\vert^{2} p_{h\vert \gamma}(h)p_{\gamma}(\gamma)dh d\gamma \right] \nonumber
	\\
	& =\log_{2}\left[1+3\rho\int  \sum_{i=1}^{N} \left[ \int \vert h_{i}\vert^{2} p_{h\vert \gamma}(h)dh \right]p_{\gamma}(\gamma) d\gamma \right]
	\label{Eqn:28}
	\end{align}
	Given the $p_{h\vert \gamma}(h)$ is Rayleigh distributed,
	\begin{align}
	& C_{UB} =\log_{2}\left[1+3\rho\int  \sum_{i=1}^{N} \sigma^{2}_{i}(\gamma) p_{\gamma}(\gamma) d\gamma \right] \nonumber \\
	& =\log_{2}\left[1+3\rho \left[ \sum_{i=1}^{N} \int \sigma^{2}_{i}(\gamma) p_{\gamma_{i}}(\gamma) d\gamma \right] \right]
	\label{Eqn:29}
	\end{align}
	we know that $\sigma^{2}_{i}(\gamma)$ is scaled Gaussian and $p_{\gamma_{i}}(\gamma)$ is the triangular distribution characterizing the AoA of the $i^{th}$ path. Substituting these values and simplifing we obtain the following form for the upper bound.
	{\scriptsize
		\begin{align}
		\nonumber
		&C_{UB}=\log_{2}\left[1+3\rho \sum_{i=1}^{N}\frac{1}{\beta^{2}_{i}}\Lambda \varsigma^2\left( \left( e^-{\frac{(\beta_{i}+\theta_{i}+\xi)^2}{\varsigma^2}}+e^-{\frac{(\beta_{i}-\theta_{i}+\xi)^2}{\varsigma^2}}\right)+\sqrt{\pi}\left(2(\xi-\theta_{i})erf\left[\frac{\theta_{i}-\xi}{\varsigma}\right]\right.\right.\right.\\ 
		&\left.\left.\left.+(\beta_{i}+\theta_{i}-\xi)erf\left[\frac{\beta_{i}+\theta_{i}-\xi}{\varsigma}\right]
		+(\beta_{i}-\theta_{i}+\xi)erf\left[\frac{\beta_{i}-\theta_{i}+\xi}{\varsigma}\right]\right)\right)\right]
		\label{Eqn:30}
		\end{align} }
	In the next section, we present simulation results for different parameter values. We also compare the upper bound to the simulated capacity of vector sensor based UWAC system.
	\begin{table}[h!]
		\caption{Simulation Parameters} 
		\centering 
		\begin{tabular}{l c c c}
			\hline \hline 
			Parameters & Symbols & Value \\[0.5ex] 
			\hline
			Range ($km$) & $R$ & 1,5,9 \\
			Water depth ($m$) & $d_w$ & 250 \\
			Transmitter depth ($m$) & $d_t$ & 150 \\
			Receiver depth ($m$) & $d_r$ & 130 \\
			Sound speed ($m/s$) & $c$ & 1520 \\
			Frequency ($kHz$) & $f$ & 5,12,22\\
			No. of beams & $Nbeams$ & 9000\\
			Water density ($kg/m^3$) & $\rho_o$ & 1027\\
			Beam take off angle & - & -90 to 90\\
			Surface type & - & Vaccum\\
			Speed at bottom (m/s)& $c_{p}$ & 1550 \\
			Bottom density (g/cm$^{3}$) & - & 1.8  \\
			Bottom attenuation (dB/$\lambda$) & - & 0.6 \\
			Ambient Noise power & - & 1.3 $\times 10^{-8}$ \\
			\hline
		\end{tabular}
		\label{bellhop_par}
	\end{table}
	\section{Numerical Results and Simulation}\label{Sec:4}
	In this section, we present the simulation results. The bound on the capacity obtained in the previous section is now analyzed to see how effective it is as a measure of performance. We simulate the capacity for vector sensor receiver case.	
	
	The simulation results show that the computed bound is tight and a good measure of performance of vector sensor based underwater communications system. We analyze the capacity of such system under different channel conditions. This study is important because as already stated earlier the capacity of underwater channel is very location specific. We vary the model parameters to simulate different channel scenarios and analyze how the capacity changes as a functions of these parameters. 
	
\begin{figure}[h!]
		\centering
		\includegraphics[width=\linewidth]{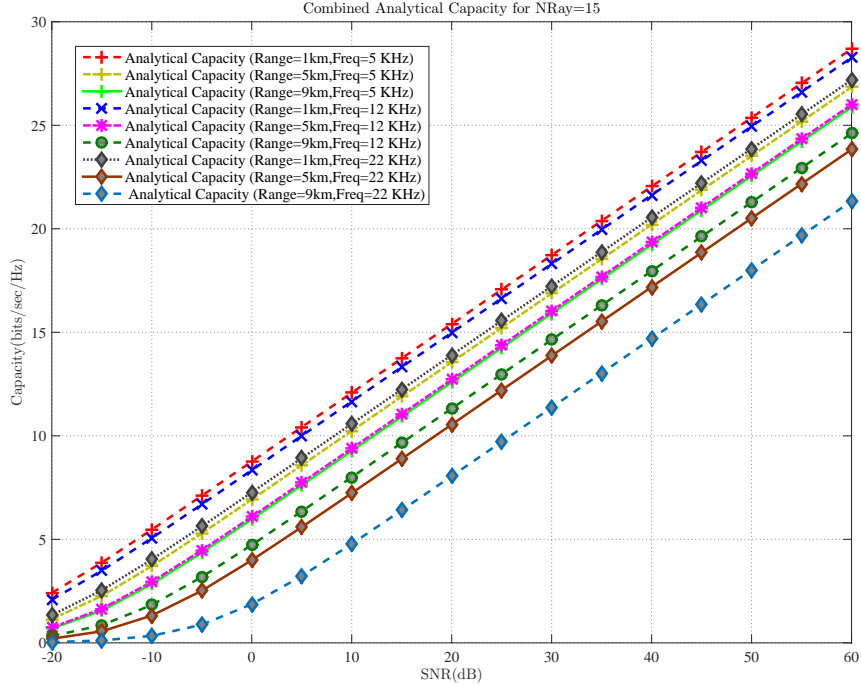}
		\caption{\label{Fig:6}{Analytical capacity as a function of SNR for different ranges and different frequencies}}
	\end{figure}

\begin{figure}[h!]
		\centering
		\includegraphics[width=\linewidth]{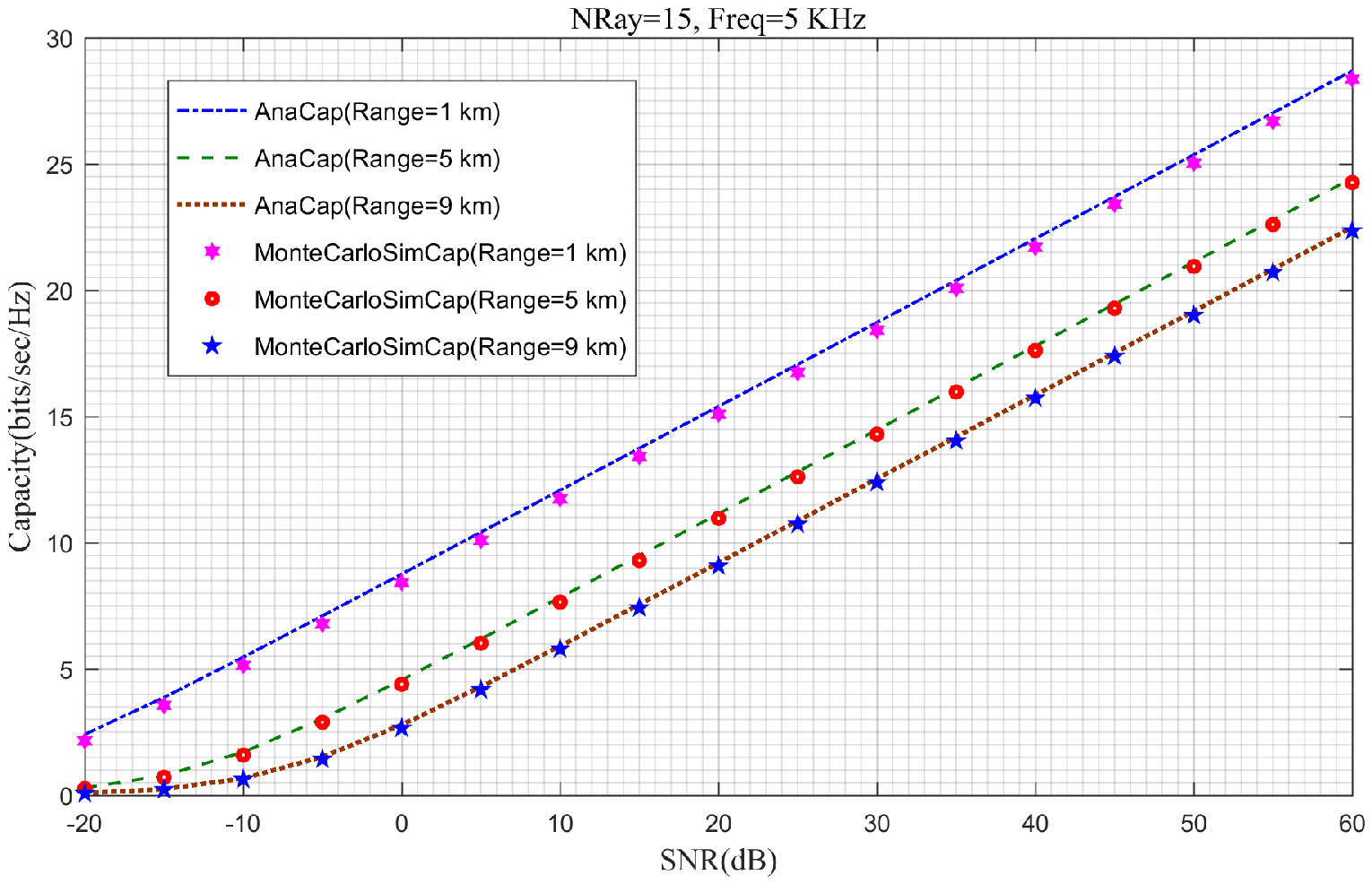}
		\caption{\label{Fig:7}{Capacity variation as a function of range for fixed number of rays (NRay=15) and frequency (f=5kHz)}}
		\vspace{-2em}
	\end{figure}

\begin{figure}[h!]
		\centering
		\includegraphics[width=\linewidth,height=3.2in]{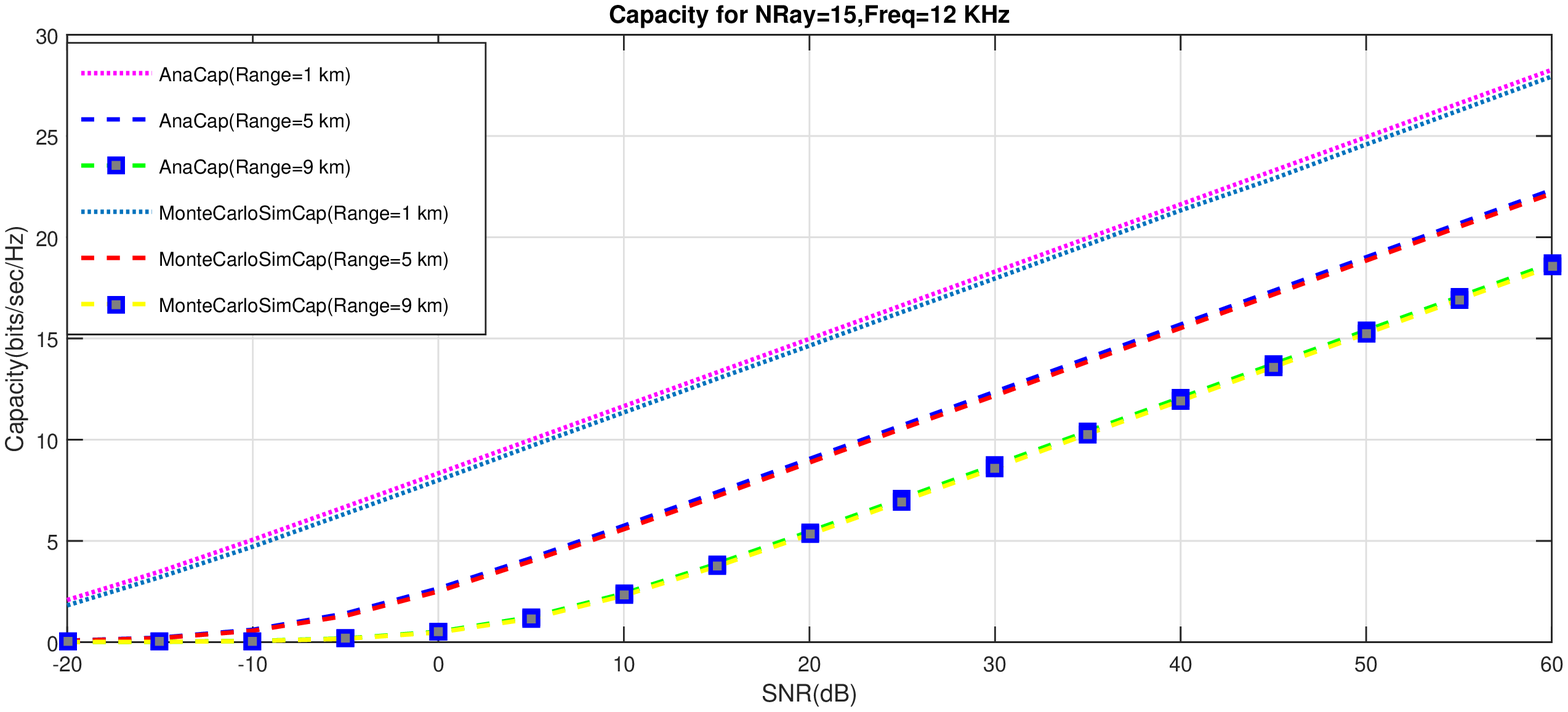}
		\caption{\label{Fig:8}{Capacity variation as a function of range for fixed number of rays (NRay=15) and frequency (f=12kHz)}}
		\includegraphics[width=\linewidth]{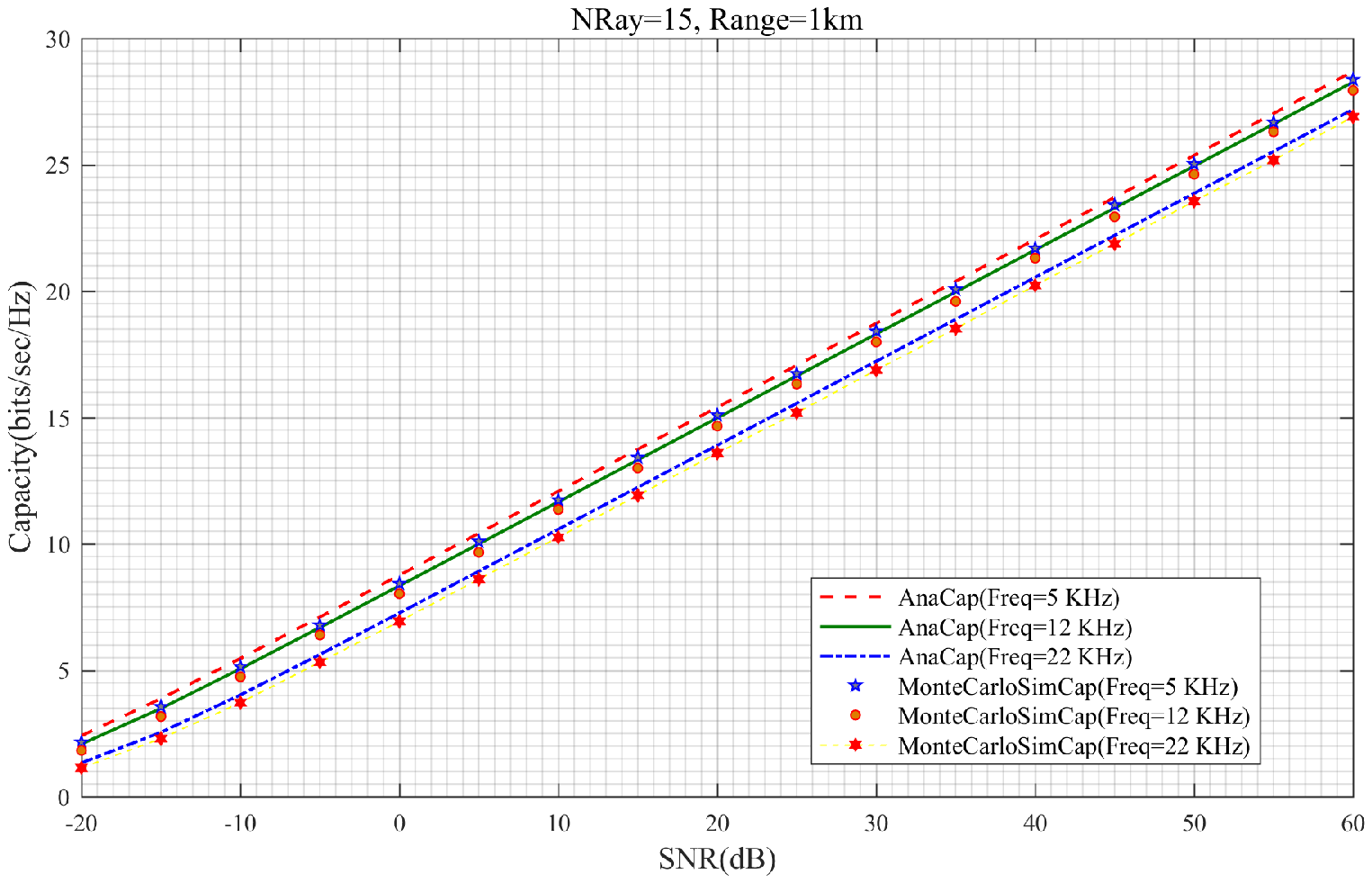}\\
		\caption{\label{Fig:9}{Capacity variation as a function of fequency for fixed number of rays (NRay=15) and range (Range=1km)}}
			\end{figure}

	Analytical capacity for different ranges with different frequency are shown in Fig.~\ref{Fig:6}. This plot provides a summary of how capacity varies as a function of these three variables viz, SNR, operating frequency and range. The capacity as a function of SNR (for a fixed frequency of operation and varying values of range) is plotted in Fig.~\ref{Fig:7}. The bound on capacity and simulated vector sensor capacity are plotted. As can be observed, the bound obtained is very tight. The experiment is repeated to obtain capacity as a function of SNR range for a fixed frequency and varying values of range refer to Fig.~\ref{Fig:8}. Again the tightness of the bound can be easily observed.
	
\begin{figure}[h!]
		\centering
		\includegraphics[width=\linewidth,height=3.2in]{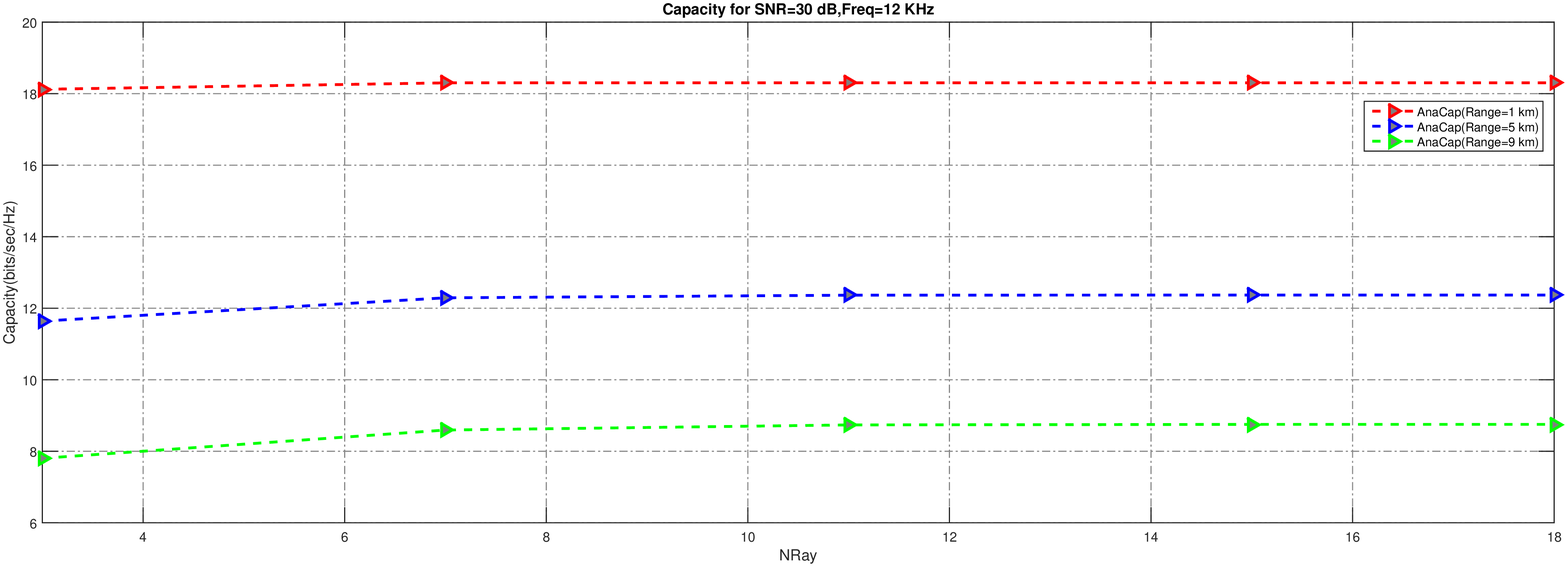}
		\caption{\label{Fig:12}{Capacity as a function of no. of rays at different ranges for fixed values of SNR and frequency.}}
	\end{figure}

\begin{figure}[h!]
		\centering
		\includegraphics[width=\linewidth,height=3.2in]{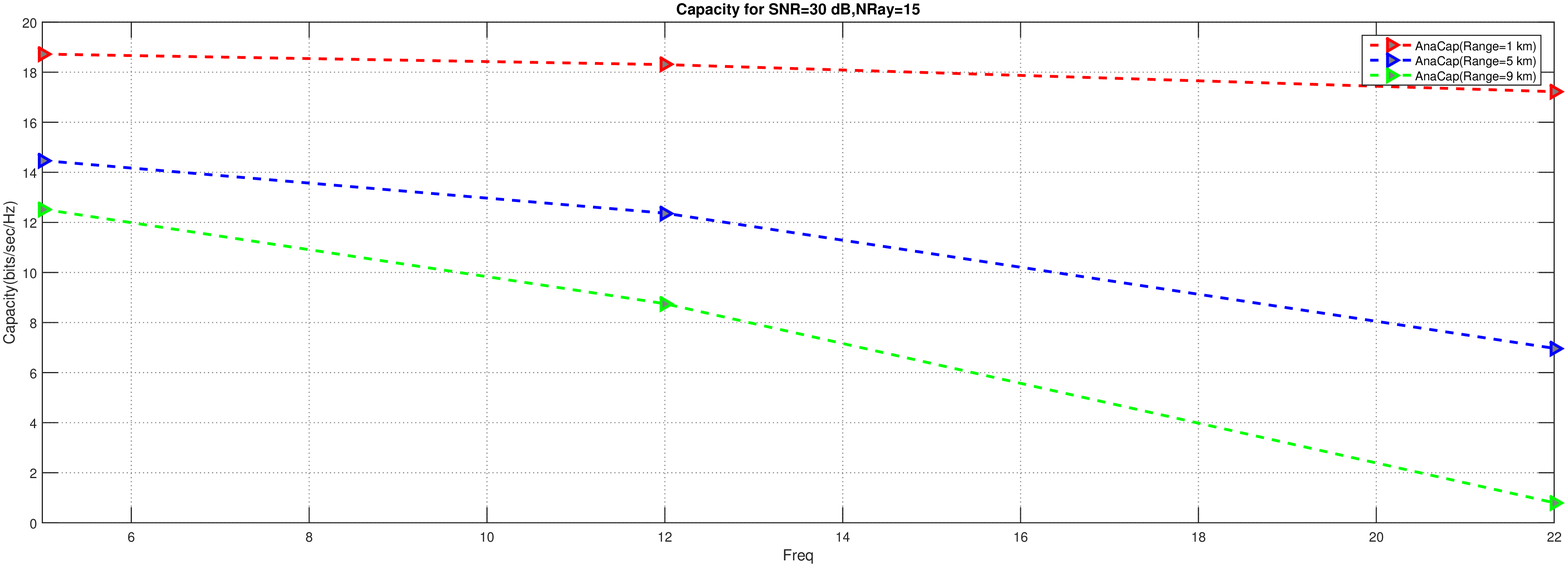}\\
		\caption{\label{Fig:13}{Capacity as a function of frequency at different ranges for fixed values of SNR and no. of rays.}}		
		\centering
		\includegraphics[width=\linewidth]{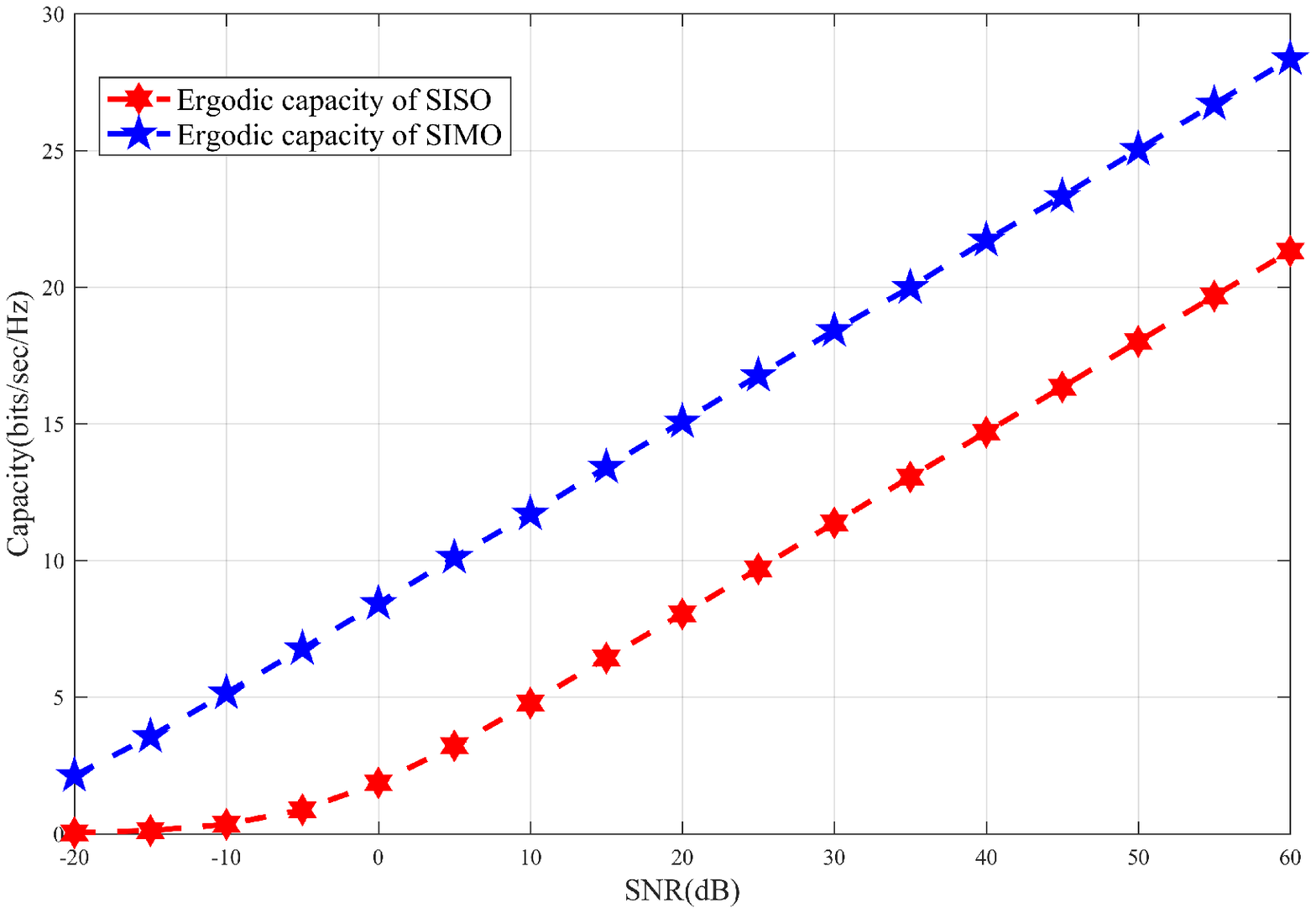}
		\caption{\label{Fig:14}{Upper bound on ergodic capacity for fixed number of rays (NRay=18) and range (Range=5km).}}
	\end{figure}

	Further experiments were performed to obtain the capacity as a function of range and that of the number of scatterers. Fig.~\ref{Fig:9}, clearly shows that capacity decreases with an increase in frequency of operation. Fig.~\ref{Fig:12} shows that as the number of scatterers increases (number of rays impinging on the receivers increases) the capacity increases and then flattens out. In Fig.~\ref{Fig:13}, we plot the capacity as a function of frequency (for fixed values of SNR and number of paths). As the range increases the plots move downwards indicating a reduction in capacity. For a fixed range the capacity decreases as a function of frequency and the decrease becomes sharper as the frequency becomes higher. Finally the plot in Fig.~\ref{Fig:14} shows that capacity of the SISO underwater communications system is higher than that of a SISO UWAC system for all values of SNR.
	\begin{figure}[h!]
	\centering
		\includegraphics[width=\linewidth]{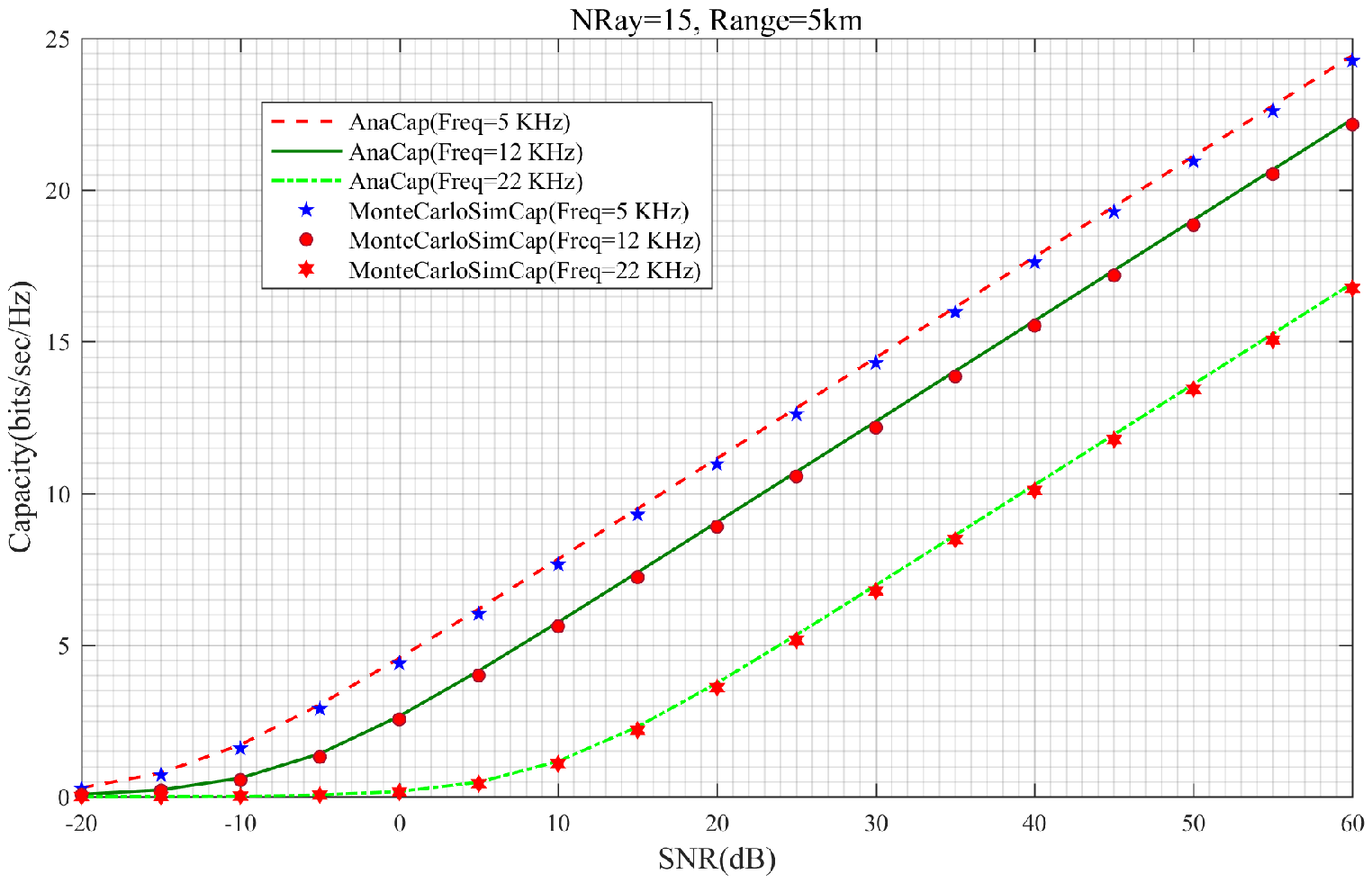}
		\caption{\label{Fig:10}{Capacity variation as a function of fequency for fixed number of rays (NRay=15) and range (Range=5km)}}
	\end{figure}
	\begin{figure}[h!]
		\includegraphics[width=\linewidth]{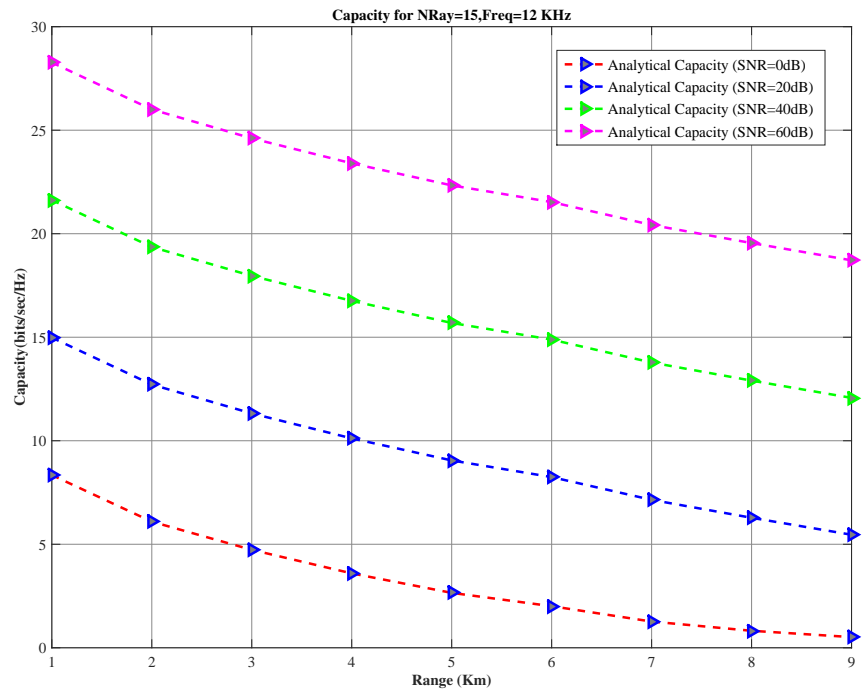}
		\caption{\label{Fig:11}{ \small Capacity as a function of range at different SNR values and fixed frequency}}
	\end{figure}
	\section{Conclusions}\label{Sec:5}
	In this paper, we derive a tight upper bound on the capacity of a vector sensor based underwater communications system. An AoA based channel representation is used to derive the results. First the AoA based framework is used to derive the density function of path gain. The same is then used to obtain a closed form expressions for the upper bound of channel capacity. The parameters of the channel have been modeled using the statistics generated using the Bellhop simulation tool. Extensive experimentation is performed and the results analyzed to check the efficacy and applicability of the proposed measure. The capacity under different channel conditions is obtained and a measure of the efficacy of vector sensor based underwater communications system is provided. The results derived are expressed in terms of parameters that represent the specific location of the channel. Thus, the result can easily be tailored for different shallow underwater channels that one encounters at different  geographical locations and at different temporal instances.
\bibliography{IET_Ref}
\bibliographystyle{plain}
\end{document}